\documentclass[aps,prd,twocolumn]{revtex4}

\usepackage{amsmath,amssymb,graphicx,color}
\usepackage{bm}
\usepackage{epsf}

\newcommand{\apjl}{ApJ}%
\newcommand{\aap}{A\&A}%
\newcommand{\araa}{ARA\&A}%
\newcommand{\mnras}{MNRAS}%
\newcommand{\pasp}{PASP}%
\newcommand{\physrep}{Phys.~Rep.}%

\newcommand{\eg}{e.g., }
\newcommand{\ie}{i.e., }
\newcommand{\Msun}{M_{\odot}}
\newcommand{\kms}{km~s$^{-1}$}

\newcommand{\Nifs}{$^{56}$Ni}

\newcommand{\Mej}{M_{\rm ej}}

\def\gsim{\mathrel{\rlap{\lower 4pt \hbox{\hskip 1pt $\sim$}}\raise 1pt
\hbox {$>$}}}
\def\lsim{\mathrel{\rlap{\lower 4pt \hbox{\hskip 1pt $\sim$}}\raise 1pt
\hbox {$<$}}}

\begin{document}

\title{Kilonova/Macronova Emission from Compact Binary Mergers}

\author{Masaomi Tanaka}
\affiliation{National Astronomical Observatory of Japan, Mitaka, Tokyo 181-8588, Japan\\ masaomi.tanaka@nao.ac.jp}

\begin{abstract}
We review current understanding of kilonova/macronova emission
from compact binary mergers
(mergers of two neutron stars or a neutron star and a black hole).
Kilonova/macronova is optical and near-infrared emission 
powered by radioactive decays of $r$-process nuclei.
Emission from the dynamical ejecta with $\sim 0.01 \Msun$
is likely to have a luminosity of $\sim 10^{40}-10^{41}\ {\rm erg\ s^{-1}}$
with a characteristic timescale of about 1 week.
The spectral peak is located in red optical or near-infrared wavelengths.
A subsequent accretion disk wind may provide an additional luminosity, 
or an earlier/bluer emission if it is not absorbed by 
the precedent dynamical ejecta.
The detection of near-infrared excess in the afterglow of short GRB 130603B
and possible optical excess in GRB 060614
supports the concept of the kilonova/macronova scenario.
At 200 Mpc distance, a typical brightness of kilonova/macronova
with $0.01 \Msun$ ejecta is expected to be about 22 mag
and the emission rapidly fades to $> 24$ mag within $\sim 10$ days 
after the merger.
Kilonova/macronova candidates can be distinguished from supernovae by
(1) the faster time evolution, (2) fainter absolute magnitudes,
and (3) redder colors.
To effectively search for such objects,
follow-up survey observations with multiple visits within $\sim 10$ days
and with multiple filters will be important.
Since the high expansion velocity ($v \sim 0.1-0.2c$) 
is a robust outcome of compact binary mergers, 
the detection of smooth spectra will be the smoking gun 
to conclusively identify the GW source.
\end{abstract}

\maketitle


\section{Introduction}
\label{sec:introduction}

Mergers of compact stars, \ie
neutron star (NS) and black hole (BH),
are promising candidates for direct detection of 
gravitational waves (GWs).
On 2015 September 14, Advanced LIGO \citep{harry10} has 
detected the first ever direct GW signals 
from a BH-BH merger (GW150914) \citep{abbott16}.
This discovery marked the dawn of GW astronomy.

NS-NS mergers and BH-NS mergers are also important
and leading candidates for the GW detection.
They are also thought to be progenitors of short-hard gamma-ray bursts
(GRBs \citep{blinnikov84,paczynski86,eichler89},
see also \citep{nakar07,berger14} for reviews).
When the designed sensitivity is realized,
Advanced LIGO \citep{harry10}, Advanced Virgo \citep{acernese15},
and KAGRA \citep{somiya12} can detect the GWs from
these events up to $\sim 200$ Mpc (for NS-NS mergers) and
$\sim 800$ Mpc (for BH-NS mergers).
Although the event rates are still uncertain,
more than one GW events per year are expected \citep{abadie10rate}.

Since localization only by the GW detectors is not accurate,
\eg more than a few 10 deg$^2$
\citep{nissanke11,ligo13,nissanke13,kelley13,abbott16review},
identification of electromagnetic (EM) counterparts is
essentially important
to study the astrophysical nature of the GW sources.
In the early observing runs of Advanced LIGO and Virgo, 
the localization accuracy can be
$> 100$ deg$^2$ \citep{kasliwal14,singer14,essick15}.
In fact, the localization for GW150914 was about 600 deg$^2$
(90 \% probability) \citep{ligo16}.

To identify the GW source from such a large localization area,
intensive transient surveys should be performed
(see \eg \citep{abbott16followup,evans16,smartt16,soares-santos16,kasliwal16,morokuma16} 
for the case of GW150914).
NS-NS mergers and BH-NS mergers are expected to emit
EM emission in various forms.
One of the most robust candidates is a short GRB.
However, the GRB may elude our detection
due to the strong relativistic beaming.
Other possible EM signals include synchrotron radio emission
by the interaction between the ejected material and interstellar gas
\citep{nakar11,piran13,hotokezaka15}
or X-ray emission from a central engine
\citep{nakamura14,metzger14magnetar,kisaka15_xray,siegel16}.

Among variety of emission mechanisms,
optical and infrared (IR) emission
powered by radioactive decay of $r$-process nuclei
\citep{li98,kulkarni05,metzger10,roberts11,goriely11,metzger12}
is of great interest.
This emission is called ``kilonova'' \citep{metzger10} 
or ``macronova'' \citep{kulkarni05}
(we use the term of kilonova in this paper).
Kilonova emission is thought to be promising:
By advancement of numerical simulations,
in particular numerical relativity \citep{shibata00,shibata05,duez10,faber12},
it has been proved that
a part of the NS material is surely ejected from NS-NS and BH-NS mergers
\citep[\eg][]{rosswog99,rosswog00,ruffert01,rosswog05,lee07,goriely11,rosswog13a,hotokezaka13,bauswein13}.
In the ejected material, $r$-process nucleosynthesis 
undoubtedly takes place
\citep[\eg][]{lattimer74,lattimer76,freiburghaus99,roberts11,goriely11,korobkin12,bauswein13,wanajo14,mendoza-temis15,just15}.
Therefore the emission powered by $r$-process nuclei is
a natural outcome from these merger events.

Observations of kilonova will also have important implications
for the origin of $r$-process elements in the Universe.
The event rate of NS-NS mergers and BH-NS mergers will be measured by
the detection of GWs.
In addition, as described in this paper, the brightness of
kilonova reflects the amount of the ejected $r$-process elements.
Therefore, by combination of GW observations and EM observations,
\ie ``multi-messenger'' observations, 
we can measure the production rate of $r$-process elements by
NS-NS and BH-NS mergers, which is essential to understand the origin
of $r$-process elements.
In fact, importance of compact binary mergers
in chemical evolution has been extensively studied in recent years
\citep{argast04,piran14,matteucci14,tsujimoto14,komiya14,cescutti15,wehmeyer15,ishimaru15,shen15,vandevoort15,hirai15}.

This paper reviews kilonova emission from compact binary mergers.
The primal aim of this paper is providing a
guide for optical and infrared follow-up observations for GW sources.
For the physical processes of compact binary mergers
and various EM emission mechanisms,
see recent reviews by Rosswog (2015) \citep{rosswog15} 
and Fern{\'a}ndez and Metzger (2016) \citep{fernandez16}.
First, we give overview of kilonova emission and
describe the expected properties of the emission 
in Section \ref{sec:kilonova}.
Then, we compare kilonova models with
currently available observations in Section \ref{sec:lessons}.
Based on the current theoretical and observational understanding,
we discuss prospects for EM follow-up observations of GW sources
in Section \ref{sec:EM}.
Finally, we give summary in Section \ref{sec:summary}.
In this paper, 
the magnitudes are given in the AB magnitude unless otherwise specified.

\section{Kilonova Emission}
\label{sec:kilonova}

\begin{figure}[t]
\begin{center}
  \includegraphics[scale=1.3]{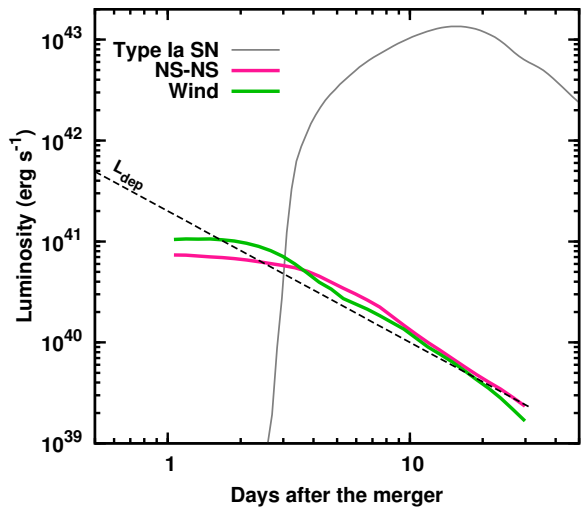}
\caption{
Bolometric light curves of a NS-NS merger model
(red, $\Mej$ = 0.01 $\Msun$ \citep{tanaka13,tanaka14}) 
and a wind model (green, $\Mej$ = 0.01 $\Msun$)
compared with a light curve of Type Ia SN model
(gray, $\Mej$ = 1.4 $\Msun$).
The black dashed line shows the deposition luminosity
by radioactive decay of $r$-process nuclei
($\epsilon_{\rm dep} = 0.5$ and $\Mej = 0.01 \Msun$).
}
\label{fig:Lbol}
\end{center}
\end{figure}

\begin{figure*}[t]
\begin{center}
\begin{tabular}{cc}
  \includegraphics[scale=1.2]{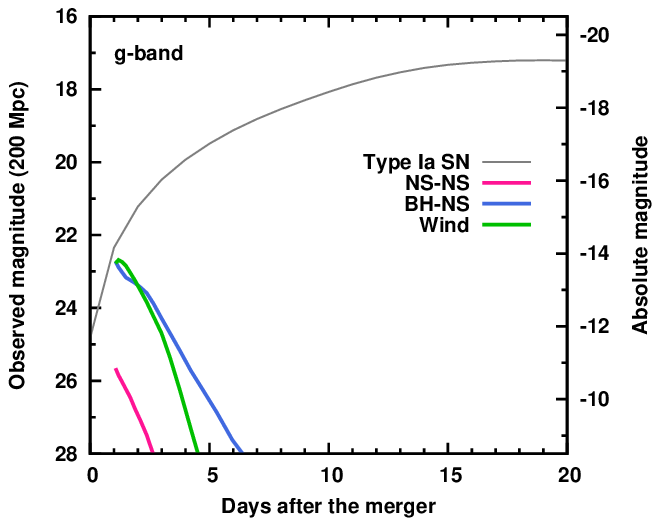} &
  \includegraphics[scale=1.2]{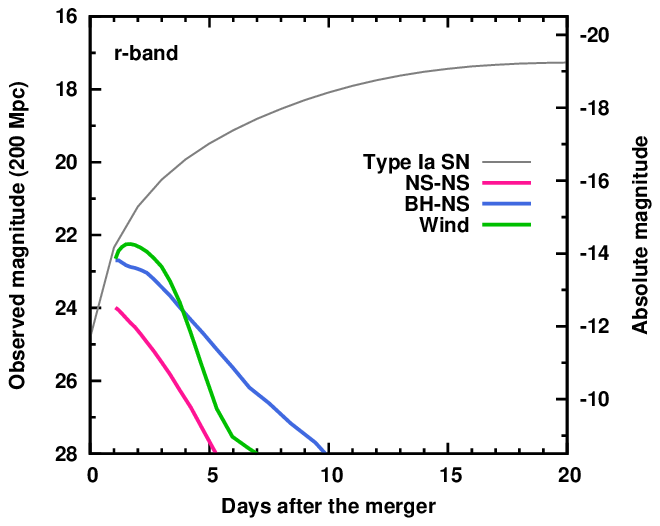} \\
  \includegraphics[scale=1.2]{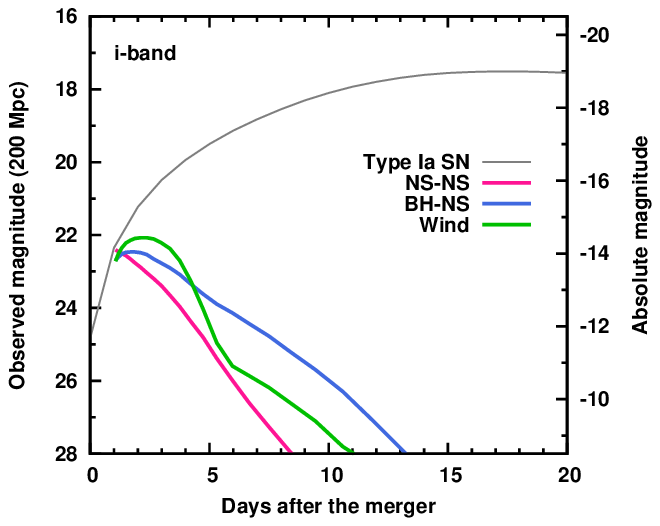} &
  \includegraphics[scale=1.2]{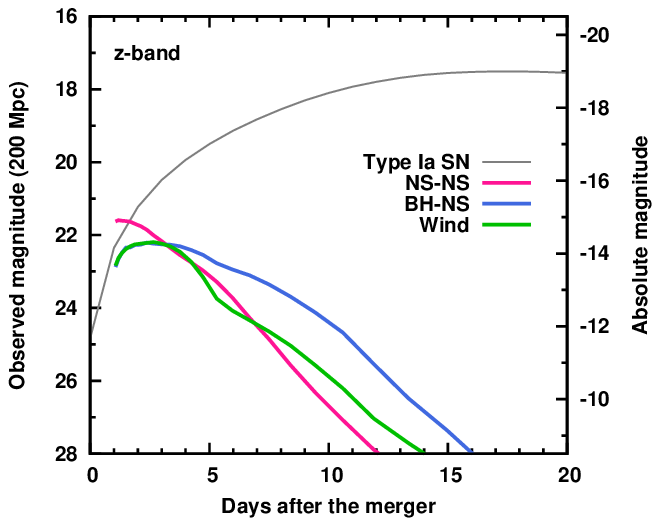} \\    
\end{tabular}
\caption{
Expected observed magnitudes of kilonova models at 200 Mpc distance
\citep{tanaka13,tanaka14}.
The red, blue, and green lines show the models of NS-NS merger (APR4-1215, \citep{hotokezaka13}),
BH-NS merger (APR4Q3a75, \citep{kyutoku13}), and a wind model (this paper),
respectively.
The ejecta mass is $\Mej = 0.01 \Msun$ for these models.
For comparison, light curve models of Type Ia SN are shown in gray.
The corresponding absolute magnitudes are indicated in the right axis.
}
\label{fig:mag}
\end{center}
\end{figure*}

\subsection{Overview}
\label{sec:overview}

The idea of kilonova emission was first introduced by 
Li \& Paczy{\'n}ski (1998) \citep{li98}.
The emission mechanism is similar to that of Type Ia supernova (SN).
The main differences are (1) 
a typical ejecta mass from compact binary mergers
is only an order of $0.01 \Msun$ ($1.4 \Msun$ for Type Ia SN),
(2) a typical expansion velocity is as high as 
$v \sim 0.1-0.2c = 30,000-60,000$ \kms ($\sim$ 10,000 \kms\ for Type Ia SN),
and (3) the heating source is decay energy of radioactive 
$r$-process nuclei (\Nifs\ for Type Ia SN).

Suppose spherical, homogeneous, and homologously expanding 
ejecta with a radioactive energy deposition.
A typical optical depth in the ejecta is $\tau = \kappa \rho R$,
where $\kappa$ is the mass absorption coefficient or ``opacity'' 
(${\rm cm^2\ g^{-1}}$),
$\rho$ is the density, and $R$ is the radius of the ejecta.
Then, the diffusion timescale in the ejecta is 
\begin{equation}
t_{\rm diff} = \frac{R}{c}\tau \simeq \frac{3 \kappa \Mej}{4 \pi cvt},
\end{equation}
by adopting $\Mej = (4\pi/3) \rho R^3$ (homogeneous ejecta) and 
$R = vt$ (homologous expansion).

When the dynamical timescale of the ejecta ($t_{\rm dyn} = R/v = t$) 
becomes comparable to the diffusion timescale, 
photons can escape from the ejecta effectively \citep{arnett82}.
From the condition of $t_{\rm diff} = t_{\rm dyn}$, 
the characteristic timescale of the emission can be written as follows:
\begin{eqnarray}
t_{\rm peak} &=& \left( \frac{3 \kappa \Mej}{4 \pi cv} \right)^{1/2} \nonumber \\
&\simeq& 8.4 \ {\rm days} \ 
\left( \frac{\Mej}{0.01 \Msun} \right)^{1/2} \nonumber \\
&& \times  \left( \frac{v}{0.1c} \right)^{-1/2}
\left( \frac{\kappa}{10\ {\rm cm^2\ g^{-1}}} \right)^{1/2}
\label{eq:tpeak}
\end{eqnarray}

The radioactive decay energy of mixture of $r$-process nuclei
is known to have a power-law dependence 
$\dot{q}(t) \simeq 2 \times 10^{10} \ {\rm erg\ s^{-1}\ g^{-1}}(t/1 \ {\rm day})^{-1.3}$
\citep{metzger10,roberts11,rosswog14,grossman14,wanajo14,lippuner15}.
By introducing a fraction of energy deposition ($\epsilon_{\rm dep}$),
the total energy deposition rate (or the deposition luminosity) is 
$L_{\rm dep} = \epsilon_{\rm dep} \Mej \dot{q}(t)$.
A majority ($\sim 90 \%$) of decay energy is released by $\beta$ decay
while the other $10 \%$ by fission \citep{metzger10}.
For the $\beta$ decay,
about $25 \%$, $25 \%$, and $50 \%$ of the energy
are carried by neutrinos, electrons, and $\gamma$-rays, respectively.
Among these, almost all the energy carried by electrons is deposited,
and a fraction of the $\gamma$-ray energy is also deposited to the ejecta.
Thus, the fraction $\epsilon_{\rm dep}$ is 
about 0.5 (see \citep{hotokezaka16} for more details).
The dashed line in Figure \ref{fig:Lbol} shows the deposition luminosity $L_{\rm dep}$ 
for $\epsilon_{\rm dep} = 0.5$ and $\Mej = 0.01 \Msun$.

Since the peak luminosity is approximated by the deposition luminosity
at $t_{\rm peak}$ (so called Arnett's law \citep{arnett82}),
the peak luminosity of kilonova can be written as follows:
\begin{eqnarray}
L_{\rm peak} &=& L_{\rm dep} (t_{\rm peak}) \nonumber \\
 &=& \epsilon_{\rm dep} \Mej \dot{q}(t_{\rm peak}) \nonumber \\
&\simeq& 1.3 \times 10^{40}\ {\rm erg\ s^{-1}} \nonumber \\
&& \times 
\left( \frac{\epsilon_{\rm dep}}{0.5} \right)^{1/2} 
\left( \frac{\Mej}{0.01 \Msun} \right)^{0.35} \nonumber \\
&& \times 
\left( \frac{v}{0.1c} \right)^{0.65}
\left( \frac{\kappa}{10\ {\rm cm^2\ g^{-1}}} \right)^{-0.65}
\label{eq:Lpeak}
\end{eqnarray}

An important factor in this analysis is the opacity 
in the ejected material from compact binary mergers.
Previously, the opacity had been assumed to be similar 
to that of Type Ia SN, \ie $\kappa \sim 0.1\ {\rm cm^2 \ g^{-1}}$ 
(bound-bound opacity of iron-peak elements).
However, recent studies \citep{kasen13,barnes13,tanaka13}
show that the opacity in the $r$-process element-rich ejecta is as high as 
$\kappa \sim 10 \ {\rm cm^2 \ g^{-1}}$
(bound-bound opacity of lanthanide elements). 
This finding largely revised our understanding of 
the emission properties of kilonova.
As evident from Eqs. (\ref{eq:tpeak}) and (\ref{eq:Lpeak}), 
a higher opacity by a factor of 100 leads to
a longer timescale by a factor of $\sim 10$, and 
a lower luminosity by a factor of $\sim 20$.

\begin{figure}[t]
\begin{center}
  \includegraphics[scale=1.2]{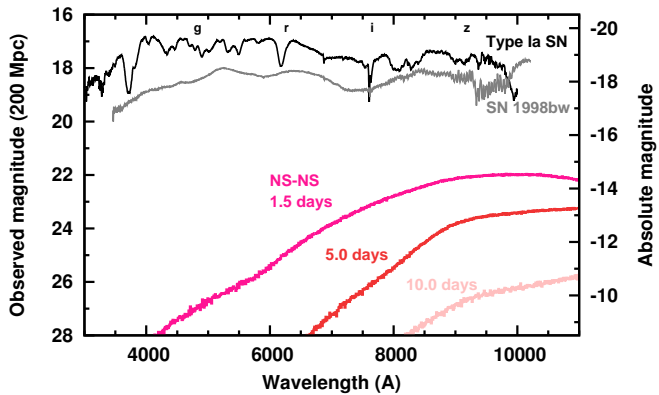}
\caption{
Expected observed spectra of the NS-NS merger model 
APR4-1215 ($\Mej = 0.01 \Msun$) 
compared with the spectra of normal Type Ia SN 2005cf
\citep{pastorello0705cf,garavini07,wang0905cf}
and broad-line Type Ic SN 1998bw \citep{galama98,iwamoto98}.
The spectra are shown in AB magnitudes ($f_{\nu}$) at 200 Mpc distance.
The corresponding absolute magnitudes are indicated in the right axis.
}
\label{fig:spec}
\end{center}
\end{figure}

\subsection{NS-NS mergers}
\label{sec:nsns}

When two NSs merge with each other,
a small part of the NSs is tidally disrupted
and ejected to the interstellar medium
(\eg \citep{rosswog99,goriely11}).
This ejecta component is mainly distributed in the orbital plane of the NSs.
In addition to this, the collision drives a strong shock,
and shock-heated material is also ejected 
in a nearly spherical manner (\eg \citep{hotokezaka13,radice16}).
As a result, NS-NS mergers have quasi-spherical ejecta.
The mass of the ejecta depends on the mass ratio and
the eccentricity of the orbit of the binary,
as well as the radius of the NS or equation of state (EOS,
\eg \citep{rosswog13,hotokezaka13,palenzuela15,sekiguchi15,radice16,sekiguchi16}):
a more uneven mass ratio and more eccentric orbit
leads to a larger amount of tidally-disrupted ejecta 
and a smaller NS radius leads to a larger amount of shock-driven ejecta.

The red line in Figure \ref{fig:Lbol} shows the expected luminosity of
a NS-NS merger model
(APR4-1215 from Hotokezaka et al. 2013 \citep{hotokezaka13}).
This model adopts a ``soft'' EOS APR4 \citep{akmal98}, 
which gives the radius of 11.1 km for a $1.35 \Msun$ NS.
The gravitational masses of two NSs are $1.2 \Msun\ + 1.5 \Msun$ and
the ejecta mass is 0.01 $\Msun$.
The light curve does not have a clear peak since
the energy deposited in the outer layer can escape earlier.
Since photons kept in the ejecta by the earlier stage 
effectively escape from the ejecta
at the characteristic timescale (Eq. \ref{eq:tpeak}),
the luminosity exceeds the energy deposition rate at
$\sim 5-8$ days after the merger.

Figure \ref{fig:mag} shows multi-color light curves of the same NS-NS merger model
(red line, see the right axis for the absolute magnitudes).
As a result of the high opacity and the low temperature \citep{kasen13},
the optical emission is greatly suppressed,
resulting in an extremely ``red'' color of the emission.
The red color is more clearly shown in Figure \ref{fig:spec},
where the spectral evolution of the NS-NS merger model
is compared with the spectra of a Type Ia SN and a broad-line Type Ic SN.
In fact, the peak of the spectrum is located at near-IR wavelengths
\citep{kasen13,barnes13,tanaka13}.

Because of the extremely high expansion velocities,
NS-NS mergers show feature-less spectra (Figure \ref{fig:spec}).
This is a big contrast to the spectra of SNe (black and gray lines),
where Doppler-shifted absorption lines of strong features can be identified.
Even broad-line Type Ic SN 1998bw (associated with long-duration GRB 980425)
show some absorption features although many lines are blended.
Since the high expansion velocity is a robust outcome of
dynamical ejecta from compact binary mergers, 
the confirmation of the smooth spectrum will be a key
to conclusively identify the GW sources.

The current wavelength-dependent radiative transfer simulations 
assume the uniform element abundances.
However, recent numerical simulations with neutrino transport show
that the element abundances in the ejecta becomes non-uniform
\citep{wanajo14,sekiguchi15,radice16,sekiguchi16}.
Because of the high temperature and neutrino absorption,
the polar region can have higher electron fractions 
($Y_e$ or number of protons per nucleon),
resulting in a wide distribution of $Y_e$ in the ejecta.
Interestingly the wide distribution of $Y_e$ is preferable for reproducing 
the solar $r$-process abundance ratios \citep{wanajo14,just15}.
This effect can have a big impact on the kilonova emission:
if the synthesis of lanthanide elements is suppressed in the polar direction, 
the opacity there can be smaller,
and thus, the emission to the polar direction can be 
more luminous with an earlier peak.

\begin{figure*}[t]
\begin{center}
  \begin{tabular}{cc}
  \includegraphics[scale=1.2]{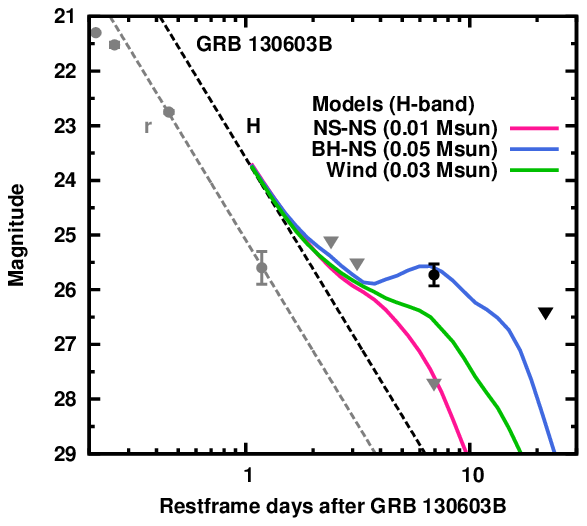} &
  \includegraphics[scale=1.2]{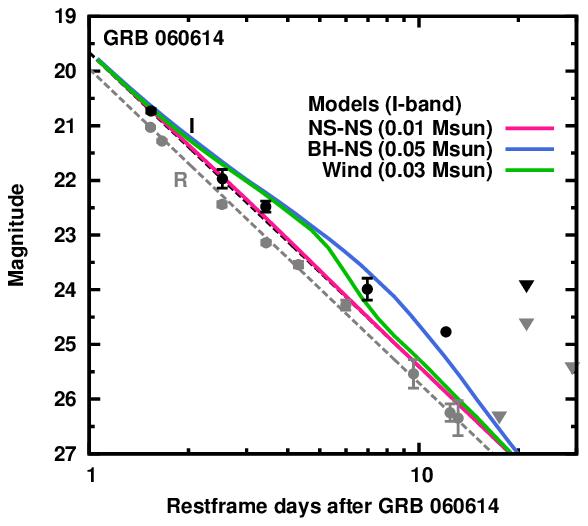} \\
    \end{tabular}
\caption{
 Comparison of kilonova models with GRB 130603B (left) and GRB 060614 (right).
 The models used in these plots are those with relatively high ejecta masses:
 APR4-1215 (NS-NS, $\Mej = 0.01 \Msun$ \citep{hotokezaka13}),
 H4Q3a75 (BH-NS, $\Mej = 0.05 \Msun$ \citep{kyutoku13}),
 and a wind model with $\Mej = 0.03 \Msun$ (this paper).
 The H4Q3a75 model is a merger of a 1.35 $\Msun$ NS
 and a 4.05 $\Msun$ BH with a spin parameter of $a=0.75$.
 This model adopts a ``stiff'' EOS H4 \citep{glendenning91,lackey06}
 which gives a 13.6 km radius for 1.35 $\Msun$ NS.
 For GRB 130603B, the afterglow component is assumed to be 
 $f_{\nu} \propto t^{-2.7}$ \citep{tanvir13,berger13}.
 For GRB 060614, it is assumed to be $f_{\nu} \propto t^{-2.3}$ 
 \citep{yang15}, which is a conservative choice (see \citep{jin15}
 for a possibility of a steeper decline).
 The observed and model magnitudes for GRB 060614 are given in 
 the Vega system as in the literature \citep{jin15}. 
}
\label{fig:obs}
\end{center}
\end{figure*}

\subsection{BH-NS mergers}
\label{sec:bhns}

Mergers of BH and NS are also important targets for GW detection
(see \citep{shibata11} for a review).
Although the event rate is rather uncertain \citep{abadie10rate},
the number of events can be comparable to that of NS-NS mergers
thanks to the stronger GW signals and thus larger horizon distances.
BH-NS mergers in various conditions have been 
extensively studied by numerical simulations
(\eg \citep{shibata06,etienne08,duez08,kyutoku10,kyutoku11}).
In particular, for a low BH/NS mass ratio (or small BH mass) and a high BH spin,
ejecta mass of BH-NS mergers can be larger than that of NS-NS mergers
\citep{kyutoku13,deaton13,foucart13,lovelace13,foucart14,kyutoku15,kawaguchi15}.
Since the tidal disruption is the dominant mechanism of the mass ejection,
a larger NS radius (or stiff EOS) gives a higher ejecta mass,
which is opposite to the situation in NS-NS mergers, 
where shock-driven ejecta dominates.

Radiative transfer simulations in BH-NS merger ejecta show
that kilonova emission from BH-NS mergers can be more luminous 
in optical wavelengths than 
that from NS-NS mergers \citep{tanaka14}.
The blues line in Figure \ref{fig:mag} show the light curve
of a BH-NS merger model (APR4Q3a75 from Kyutoku et al. 2013 \citep{kyutoku13}),
a merger of a 1.35 $\Msun$ NS and a 4.05 $\Msun$ BH
with a spin parameter of $a=0.75$.
The mass of ejecta is $\Mej = 0.01 \Msun$.
Since BH-NS merger ejecta are highly anisotropic 
and confined to a small solid angle, 
the temperature of the ejecta can be higher for a given mass of the ejecta,
and thus, the emission tends to be bluer than in NS-NS mergers.
Therefore, even if the bolometric luminosity is similar,
the optical luminosity of BH-NS mergers can be higher than 
that of NS-NS mergers.

It is emphasized that the mass ejection from BH-NS mergers has
a much larger diversity compared with NS-NS mergers,
depending on the mass ratio, the BH spin, and its orientation.
As a result, the expected brightness also has a large diversity.
See Kawaguchi et al. (2016) \citep{kawaguchi16} for 
the expected kilonova brightness for a wide parameter space.

\subsection{Wind components}
\label{sec:wind}

After the merger of two NSs,
a hypermassive NS is formed at the center, and 
it subsequently collapses to a BH.
During this process, accretion disk surrounding 
the central remnant is formed.
A BH-accretion disk system is also formed in BH-NS mergers.
From such accretion-disk systems, 
an outflow or disk ``wind'' can be driven by neutrino heating, 
viscous heating, or nuclear recombination
\citep{dessart09,fernandez13,perego14,kiuchi14,kiuchi15,fernandez15nsns,fernandez15bhns,just15}.
A typical velocity of the wind is $v=10,000-20,000$ \kms, 
slower than the precedent dynamical ejecta.
Although the ejecta mass largely depends on the ejection mechanism,
a typical mass is likely an order of $\Mej = 0.01 \Msun$ or even larger.

This wind component is another important source of kilonova emission
\citep{fernandez13,metzger14,perego14,kasen15,martin15}.
The emission properties depend on the element composition in the ejecta.
In particular, if a high electron fraction ($Y_e \gsim 0.25$) is realized by 
the neutrino emission from a long-lived hypermassive NS
\citep{metzger14,kasen15} or shock heating in the outflow \citep{kiuchi15}, 
synthesis of lanthanide elements can be suppressed in the wind.
Then, the resulting emission can be bluer 
than the emission from the dynamical ejecta
thanks to the lower opacity \citep{kasen13,tanaka13}.
This component can be called as ``blue kilonova'' \citep{fernandez16}.

To demonstrate the effect of the low opacity, we show 
a simple wind model in Figures \ref{fig:Lbol} and \ref{fig:mag}.
In this model, we adopt a spherical ejecta of $\Mej = 0.01 \Msun$
with a density structure of $\rho \propto r^{-2}$ 
from $v = 0.01c$ to $0.1c$ (with the average velocity of $v \sim 20,000$ \kms).
The elements in the ejecta are assumed to be lanthanide-free:
only the elements of $Z=31-54$ are included 
with the solar abundance ratios. 
As shown by previous works \citep{kasen15}, 
the emission from the such a wind can peak earlier than
that from the dynamical ejecta
(Figure \ref{fig:Lbol}) and the emission is bluer 
(Figure \ref{fig:mag}).

Note that this simple model neglects the presence of the 
dynamical ejecta outside of the wind component.
The effect of the dynamical ejecta is in fact important,
because it works as a ``lanthanide curtain'' \citep{kasen15}
absorbing the emission from the disk wind.
Interestingly, as described in Section \ref{sec:nsns},
the polar region of the dynamical ejecta can have a higher $Y_e$,
and the ``lanthanide curtain'' may not be present in the direction.
Also, in BH-NS mergers, the dynamical ejecta is distributed 
in the orbital plane, and disk wind can be directly observed
from most of the lines of sight.
If the wind component is dominant for kilonova emission,
and can be directly observed,
the spectra are not as smooth as the spectra of dynamical ejecta 
because of the slower expansion \citep{kasen15}.
More realistic simulations capturing all of these situations 
will be important to understand the emission from the disk wind.

\begin{figure}[t]
\begin{center}
   \includegraphics[scale=1.6]{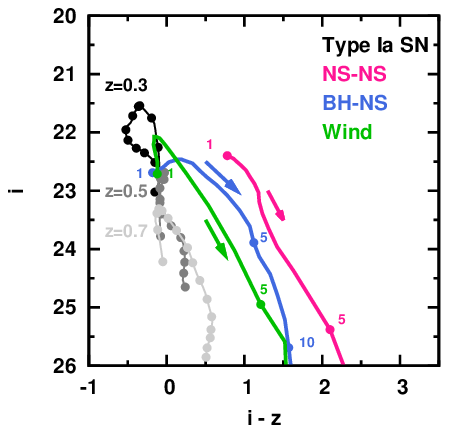}
    \includegraphics[scale=1.6]{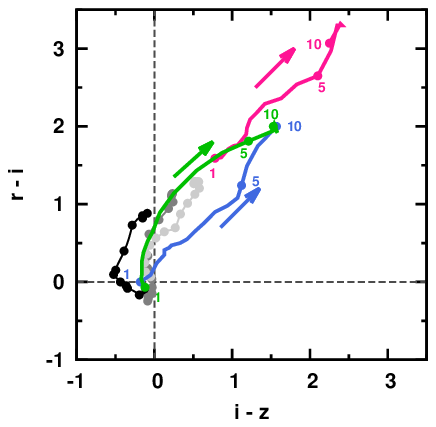} 
  \caption{
Color-magnitude diagram (top) and color-color diagram (bottom)
for compact binary merger models ($\Mej = 0.01 \Msun$) at 200 Mpc
compared with Type Ia SN with similar observed magnitudes 
($z=$ 0.3, 0.5, and 0.7).
For Type Ia SN, we use spectral templates \citep{nugent02}
with $K$-correction.
The numbers for binary merger models show time from the merger in days
while dots for Type Ia SN are given with 5-day interval. 
}
\label{fig:color}
\end{center}
\end{figure}

\section{Lessons from Observations}
\label{sec:lessons}

Since short GRBs are believed to be
driven by NS-NS mergers or BH-NS mergers (see \eg \citep{nakar07,berger14}),
models of kilonova can be tested by the observations of short GRBs.
As well known, SN component has been detected in
the afterglow of long GRBs (see \citep{woosley06,cano16} for reviews).
If kilonova emission occurs,
the emission can be in principle visible on top of the afterglow,
but such an emission had eluded the detection for long time \citep{kann11}.

In 2013, a clear excess emission was detected
in the near-IR afterglow of GRB 130603B \citep{berger13,tanvir13}.
Interestingly, the excess was not visible in the optical data.
Since this behavior nicely agrees with the expected properties of kilonova,
the excess is interpreted to be the kilonova emission.

The left panel of Figure \ref{fig:obs} shows kilonova models compared with
the observations of GRB 130603B.
The observed brightness of the near-IR excess in GRB 130603B
requires a relatively large ejecta mass of
$\Mej \gsim 0.02 \Msun$ \citep{berger13,tanvir13,hotokezaka13b,piran14}.
As pointed out by Hotokezaka et al. (2013) \citep{hotokezaka13b},
this favors a soft EOS for a NS-NS merger model (\ie more shock-driven ejection)
while a stiff EOS for a BH-NS merger model (\ie more tidally-driven ejection).
Another possibility to explain the brightness may be an additional 
emission from the disk wind
(green line in Figure \ref{fig:obs}, see \citep{metzger14,kasen15}).

Note that the excess was detected only at one epoch in one filter.
Therefore, other interpretations are also possible,
\eg emission by the external shock \citep{jin13}
or by a central magnetar \citep{yu13,fan13}, 
or thermal emission from newly formed dust \citep{takami14}.
Importantly, a late-time excess is also visible in X-ray \citep{fong14},
and thus, the near-IR and X-ray excesses might be caused by
the same mechanism, possibly the central engine
\citep{kisaka15a,kisaka16}.

Another interesting case is GRB 060614.
This GRB was formally classified as a long GRB because the duration
is about 100 sec.
However, since no bright SN was accompanied,
the origin was not clear \citep{gehrels06,fynbo06,dellavalle06,gal-yam06}.
Recently the existence of a possible excess in the optical
afterglow was reported \citep{yang15,jin15}.
The right panel of Figure \ref{fig:obs} shows the comparison between
GRB 060614 and the same sets of the models.
If this excess is caused by kilonova,
a large ejecta mass of $\Mej \sim 0.1 \Msun$ is required.
This fact may favor a BH-NS merger scenario with
a stiff EOS \citep{yang15,jin15}.
It is however important to note that
the emission from BH-NS merger has a large variation,
and such an effective mass ejection requires
a low BH/NS mass ratio and a high BH spin \citep{kawaguchi16}.
See also \citep{jin16} for possible optical excess
in GRB 050709, a genuine short GRB with a duration of 0.5 sec
\citep{villasenor05,Hjorth05Nature,fox05,covino06}.
If the excess is attributed to kilonova,
the required ejecta mass is $\Mej \sim 0.05 \Msun$.

Finally, an early brightening in optical data of GRB 080503 at $t \sim 1-5$ days
can also be attributed to kilonova \citep{perley09}
although the redshift of this object is unfortunately unknown.
Kasen et al. (2015) \citep{kasen15} give a possible interpretation
with the disk wind model.
Note that a long-lasting X-ray emission was
also detected in GRB 080503 at $t \lsim 2$ days, and
it may favor a common emission mechanism for optical and X-ray
\citep{gao15,kisaka16}.

\section{Prospects for EM Follow-up Observations of GW sources}
\label{sec:EM}

Figure \ref{fig:mag} shows the expected brightness of 
compact binary merger models at 200 Mpc (left axis).
All the models assume a canonical ejecta mass of $\Mej = 0.01 \Msun$, 
and therefore, the emission can be brighter or fainter depending 
on the merger parameters and the EOS (see Section \ref{sec:kilonova}).
Keeping this caveat in mind, 
typical models suggest that the expected kilonova brightness at 200 Mpc
is about 22 mag in red optical wavelengths ($i$ or $z$ bands)
at $t<5$ days after the merger.
The brightness quickly declines to $> 24$ mag 
within $t \sim 10$ days after the merger.
To detect this emission, we ultimately need 8m-class telescopes.
Currently the wide-field capability for 8m-class telescopes
is available only at the 8.2m Subaru telescope:
Subaru/Hyper Suprime-Cam (HSC) has
the field of view (FOV) of 1.77 deg$^2$ \citep{miyazaki06,miyazaki12}.
In future, the 8.4m Large Synoptic Survey Telescope 
(LSST) with 9.6 deg$^2$ FOV will be online \citep{ivezic08,lsst09}.
Note that targeted galaxy surveys are also effective 
to search for the transients associated with galaxies \citep{gehrels16,singer16}.

It is again emphasized that the expected brightness of kilonova 
can have a large variety.
If the kilonova candidates seen in 
GRB 130603B ($\Mej \gsim 0.02 \Msun$)
and GRB 060614 ($\Mej \sim 0.1 \Msun$)
are typical cases (see Section \ref{sec:lessons}),
the emission can be brighter by $\sim$ 1-2 mag.
In addition, there are also possibilities of bright, precursor emission
(\eg \citep{metzger14magnetar,metzger15,kisaka15a})
which are not discussed in depth in this paper.
And, of course, the emission is brighter for objects at shorter distances.
Therefore, surveys with small-aperture telescopes 
(typically with wider FOVs) are also important.
See \eg Nissanke et al. (2013) \citep{nissanke13} and 
Kasliwal \& Nissanke (2014) \citep{kasliwal14} for 
detailed survey simulations for various expected brightness
of the EM counterpart.

A big challenge for identification of the GW source is 
contamination of SNe.
NS-NS mergers and BH-NS mergers are rare events compared with SNe,
and thus, much larger number of SNe are detected 
when optical surveys are performed over $10$ deg$^2$
(see \citep{smartt16,soares-santos16,kasliwal16} for the case of GW 150914).
Therefore, it is extremely 
important to effectively select the candidates of kilonova
from a larger number of SNe.

To help the classification, 
color-magnitude and color-color diagrams for 
the kilonova models and Type Ia SNe are shown in Figure \ref{fig:color}.
The numbers attached with the models are days after the merger
while dots for SNe are given with 5-day interval.
According to the current understanding,
the light curves of kilonova can be characterized as follows:
\begin{enumerate}
\item 
The timescale of variability should be shorter than that of SNe
(Figure \ref{fig:mag}).
This is robust since the ejecta mass from compact binary mergers
is much smaller than SNe.

\item 
The emission is fainter than SNe.
This is also robust because of the smaller ejecta mass and thus
the lower available radioactive energy (Figure \ref{fig:Lbol}).

\item
The emission are expected to be redder than SNe.
This is an outcome of a high opacity in the ejecta,
but the exact color depends on the ejecta composition 
(\citep{kasen13,tanaka14,metzger14,kasen15}, Section \ref{sec:kilonova}).
\end{enumerate}

Therefore, in order to effectively search for 
the EM counterpart of the GW source,
multiple visits in a timescale of $<10$ days will be important
so that the rapid time evolution can be captured.
Surveys with multiple filters are also helpful to use color information.
As shown in Figure \ref{fig:color},
observed magnitudes of kilonovae at $\sim 200$ Mpc
are similar to those of SNe at larger distances ($z \gsim 0.3$ for Type Ia SNe).
Therefore, if redshifts of the host galaxies are estimated, 
kilonova candidates can be further selected
by the close distances and the intrinsic faintness.

\section{Summary}
\label{sec:summary}

The direct detection of GWs from GW 150914 opened GW astronomy.
To study the astrophysical nature of the GW sources,
the identification of the EM counterparts is essentially important.
In this paper, we reviewed the current understanding of 
kilonova emission from compact binary mergers.

Kilonova emission from the dynamical ejecta of 0.01 $\Msun$
has a typical luminosity is an order of $10^{40}-10^{41}\ {\rm erg\ s^{-1}}$
with the characteristic timescale of about 1 week.
Because of the high opacity and the low temperature, 
the spectral peak is located at red optical or near-IR wavelengths.
In addition to the emission from the dynamical ejecta,
a subsequent disk wind can cause an additional emission 
which may peak earlier with a bluer color
if the emission is not absorbed by the precedent ejecta.

The detection of excess in GRB 130603B (and possibly GRB 060614)
supports the kilonova scenario.
If the excesses found in these objects are attributed 
to the kilonova emission,
the required ejecta masses are $\Mej \gsim 0.02 \Msun$, 
and $\Mej \sim 0.1 \Msun$, respectively.
The comparison between such observations and numerical simulations
gives important insight to study the progenitor of compact binary
mergers and EOS of NS.

At 200 Mpc distance, a typical peak brightness of kilonova emission is 
about 22 mag in the red optical wavelengths ($i$ or $z$ bands).
The emission quickly fades to $> 24$ mag within $\sim 10$ days.
To distinguish GW sources from SNe, 
observations with multiple visits in a timescale of $<10$ days 
are important to select the objects with rapid temporal evolution.
The use of multiple filters are also helpful to select red objects.
Since the extremely high expansion velocities ($v \sim 0.1-0.2c$) are 
unique features of dynamical mass ejection from compact binary mergers,
detection of extremely smooth spectrum will 
be the smoking gun to conclusively identify the GW sources.

\vspace{1cm}

\section*{Conflict of Interests}
The author declares that there is no conflict of interests regarding the publication of this paper.

\section*{Acknowledgments}
The author thanks Kenta Hotokezaka, Yuichiro Sekiguchi,
Masaru Shibata, Kenta Kiuchi, Shinya Wanajo, Koutarou Kyutoku,
Kyohei Kawaguchi, Keiichi Maeda, Takaya Nozawa, and Yutaka Hirai
for fruitful discussion on compact binary mergers,
nucleosynthesis, and kilonova emission.
The author also thanks Nozomu Tominaga, Tomoki Morokuma, 
Michitoshi Yoshida, Kouji Ohta, and the J-GEM collaboration
for valuable discussion on EM follow-up observations.
Numerical simulations presented in this paper 
were carried out with Cray XC30 at Center for Computational Astrophysics, 
National Astronomical Observatory of Japan.
This research has been supported 
by the Grant-in-Aid for Scientific Research of
the Japan Society for the Promotion of Science (24740117, 15H02075)
and Grant-in-Aid for Scientific Research on Innovative Areas
of the Ministry of Education, Culture, Sports, Science and Technology 
(25103515, 15H00788).


\end{document}